%% file: main.tex
\documentclass[12pt]{article}

\usepackage{scicite}
\usepackage{graphicx}
\usepackage{siunitx}
\usepackage{times}
\usepackage{lineno}
\usepackage{caption}
\usepackage{subcaption}
\usepackage{xcolor}
\usepackage[normalem]{ulem}

\usepackage{amsmath}
\usepackage{amssymb}

\newcommand{\AuAu}{Au+Au}
\newcommand{\RuRu}{Ru+Ru}
\newcommand{\ZrZr}{Zr+Zr}

\newcommand{\sqrtsNN}{\ensuremath{\sqrt{s_{_\mathrm {NN}}}}}
\newcommand{\npart}{\ensuremath{\langle N_{\mathrm{part}}\rangle}}

\newcommand{\pT}{\ensuremath{p_\mathrm{T}}}
\newcommand{\mT}{\ensuremath{m_\mathrm{T}}}
\newcommand{\gev}{GeV/$c$}
\newcommand{\address}[1]{\vspace{0pt}{\small #1}\\}

\newenvironment{sciabstract}{%
\begin{quote} \bf}
{\end{quote}}

\title{Tracking the baryon number with nuclear collisions\\
{\normalsize{ Short Title: Do quarks or gluons carry baryon number?}} }
\date{}

\begin{document}
\maketitle

\include{author.tex}


\begin{sciabstract}
Baryon quantum number is found to be conserved since baryogenesis in the early Universe. Conventionally, each fractionally charged valence quark is understood to carry one-third of a baryon number. An alternative hypothesis posits that baryon number is instead carried by a baryon junction -- a non-perturbative, Y-shaped gluonic configuration.
Neither scenario has been verified experimentally. The STAR Collaboration reports measurements at mid-rapidity of baryon number ($\boldsymbol{B}$) over the electric charge number difference ($\boldsymbol{\Delta Q}$) in isobar nuclear collisions, and the net-proton yield along rapidity in photonuclear collisions. A larger $\boldsymbol{B/\Delta Q}$ ratio and less asymmetric net-proton yield are observed than predicted from models assigning baryon number to valence quarks. These findings, corroborated by previous measurements  in Au+Au collisions, disfavor the valence quark picture.
\end{sciabstract}

A handful of fundamental conservation laws in physics govern the evolution of the existing world. Among the associated quantities, baryon number ($B$) has been found to be conserved since baryogenesis, shortly after the birth of the Universe ~\cite{Sakharov:1967dj, Sarkar:2008xir}. It is associated with hadrons containing three quarks, such as protons and neutrons, and dictates the baryon asymmetry in the present Universe. The Standard Model provides a framework for understanding elementary particles, building blocks of our visible Universe, and interactions between them. These elementary particles possess intrinsic properties such as electric charge ($Q$), mass, and spin. One subset of these elements, the quarks, also carry color charge and are subject to the strong interaction mediated by gluons, as described by Quantum Chromodynamics (QCD). 

\begin{figure}[ht!]
    \includegraphics[width=\linewidth]{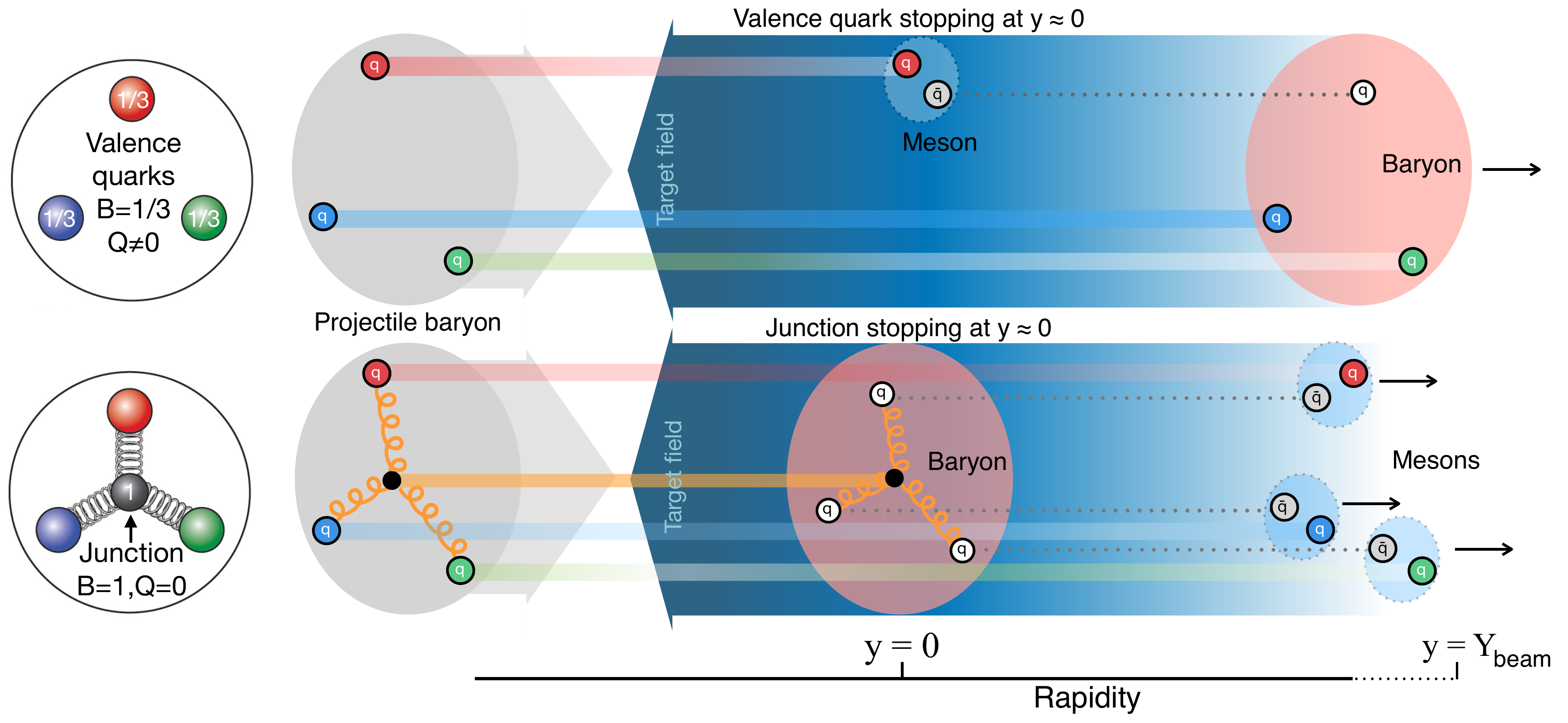}
    \caption{{\bf Baryon number carrier and transport mechanism.} Left: illustration of two possible carriers of the baryon number (top: valence quarks; bottom: baryon junction).  Right: most likely outcome of a single scattering interaction that can occur for the two scenarios: valence-quark picture (top), which yields a mid-rapidity meson (dashed blue circle) and a forward baryon; or junction picture (bottom), which produces a mid-rapidity baryon and forward mesons. Solid colored circles represent the original valence quarks, whereas open circles denote quark–antiquark pairs produced from the vacuum. The “projectile” denotes the nucleus explicitly drawn with either valence quarks or a baryon junction, whereas the “target” refers to the other colliding nucleus. 
 }   
    \label{fig:cartoon}
\end{figure}

Conventionally, quarks are believed to carry a baryon number of $B=1/3$ \cite{Pire:2021hbl}. Therefore, three quarks combine to form a baryon with $B=1$, whereas a quark-antiquark pair form a meson with $B=0$. An alternative picture, proposed in the 1970s \cite{Artru:1974zn,Rossi:1977cy}, assigns the baryon number not to the quarks individually but to a Y‑shaped gluonic configuration known as the baryon junction. In this framework, the junction is the topological object that carries and transports the conserved baryon number. Its three gluonic strings connect to the valence quarks, and the motion of the junction through spacetime defines the flow of baryon number~\cite{Kharzeev:1996sq}. 
Such a junction structure obeys the fundamental physics conservation laws and does not contradict any existing observations~\cite{Rossi:1977cy,Kharzeev:1996sq}. Although this topological object cannot be obtained from perturbative quark and gluon interactions, it has been studied non-perturbatively in Lattice QCD \cite{Suganuma:2004zx,Takahashi:2000te,Bissey:PhysRevD.76.114512,takahashi2002detailed,ichie2003flux,bornyakov2004baryonic,suganuma2022quantum}. 
The top-left panel of Fig.~\ref{fig:cartoon} illustrates a baryon with three valence quarks each carrying $B=1/3$ and $Q\neq0$, whereas the bottom-left panel illustrates the alternative picture of a baryon junction carrying $B=1$ and $Q=0$. The three ends of the baryon junction are connected to valence quarks, making it indistinguishable for most physical processes whether the junction or valence quarks carry the baryon number~\cite{Kharzeev:1996sq,Csorgo:2001sq}. For this reason, neither scenario has been unequivocally verified experimentally. Note that junction connections without valence quarks have also been predicted, including possible configurations of baryonium tetraquarks~\cite{Rossi:2016szw}, exotic gluonic states of junction and anti-junction such as baryonium glueballs, gluonic graphene and buckyballs~\cite{Csorgo:2001sq}. These exotic configurations await experimental confirmation. 

The phenomenon of ``baryon transport" in high-energy nuclear collisions, referring to the way the baryon number is moved around in the momentum space, provides a potential tool to identify the carrier of the baryon number~\cite{Lewis:2022arg,Kharzeev:1996sq,Gyulassy:1997mz}. In these collisions, significant excesses of baryons over antibaryons have been observed at mid-rapidity ($y\sim0$) in the center-of-mass frame; here rapidity ($y$) is defined as
\begin{equation}
    y=\frac{1}{2}\ln\frac{E+p_{z}c}{E-p_{z}c},
\end{equation} where $E$ and $p_{z}$ are the energy and the $z$ component of momentum along beam direction for a particle whereas $c$ is the speed of light~\cite{Gyulassy:1997mz,NA49:1998gaz,BRAHMS:2009wlg,STAR:2008med}. It can be understood as a scaled velocity along the $z$-axis that is convenient for relativistic calculations. Owing to baryon number conservation, these excess baryons must contain the baryon numbers that get transported from the colliding nuclei at beam rapidity ($y=\pm Y_{\rm beam}$) to mid-rapidity. The right panels of Fig.~\ref{fig:cartoon} illustrate how the baryon number is likely to be transported in a single scattering for either valence quarks (top) or baryon junctions (bottom). In the valence quark picture, if a valence quark is struck, a meson is likely produced at mid-rapidity and the baryon continues close to $Y_{\rm beam}$, whereas in the junction picture, a junction can be scattered to mid-rapidity and coupled with quarks from quark-antiquark pairs popping out of vacuum to form baryons. In the latter case, valence quarks are likely to form mesons near $Y_{\rm beam}$. A key difference emerges between the two mechanisms, namely that valence quarks, which carry a large fraction of the colliding baryon's momentum, are difficult to transport all together to form baryons at $y\sim0$, whereas baryon junctions, composed of low-momentum gluons, are easier to be transported to mid-rapidity. Therefore, one expects enhanced baryon transport for the junction picture compared to that of the valence quark picture. 

The STAR Collaboration at the Relativistic Heavy Ion Collider (RHIC) presents three independent approaches to investigate the question of what carries the baryon number~\cite{Lewis:2022arg}. First, we measure the rapidity dependence of proton, antiproton and net-proton (proton minus antiproton) yields in photon-gold ($\gamma$+Au) interactions~\cite{Artru:1990wq}. Second, we point out that previously measured net-proton yields at mid-rapidity from Au+Au collisions of different energies~\cite{STAR:2008med,STAR:2017sal} show a striking exponential scaling as a function of the rapidity shift from $Y_{\rm{beam}}$ to $y\sim0$~\cite{Lewis:2022arg}. Third, we study charge and baryon transport to mid-rapidity, utilizing the dataset from isobar ${}^{96}_{44}\rm{Ru}$+${}^{96}_{44}\rm{Ru}$ and ${}^{96}_{40}\rm{Zr}$+${}^{96}_{40}\rm{Zr}$ collisions~\cite{STAR:2021mii}. These results are compared with theoretical expectations based on valence quarks or baryon junction scenarios.

\section*{Baryon transport along rapidity} \label{Sec:GammaA}
Photon-induced interactions~\cite{Artru:1990wq} on nuclei provide a clean environment to study the baryon transport, as the incoming photon does not carry any baryon number. In such interactions, the rapidity dependence of the net-baryon yield is expected to follow an exponential distribution~\cite{Kharzeev:1996sq}:
\begin{linenomath*}
\begin{equation}
f(y)\propto \exp(-\alpha_B\Delta y),
\label{eq:ydep}
\end{equation}
\end{linenomath*}
where $\Delta y=Y_{\mathrm{beam}}-y$. For a given beam energy ($Y_{\mathrm{beam}}$), the rapidity dependence becomes $f(y) \propto \exp(\alpha_B y)$. The slope parameter ($\alpha_B$) is predicted to be within $0.42 < \alpha_B < 1$ by the baryon junction framework based on the Regge theory~\cite{Kharzeev:1996sq}. The upper and lower limits correspond to junction-junction (J + J) and junction-Pomeron (J + $\mathbb{P}$) interactions, where a Pomeron is a theorized two-gluon configuration.  In the Regge theory, interactions among junctions and Pomerons are responsible for the increase of the elastic and inelastic cross sections in $p$+$p$ and $p$+anti-$p$ collisions with beam energy, from which the rapidity dependence of baryon transport could be inferred~\cite{Kharzeev:1996sq}. This predicted $\alpha_B$ range also applies to baryon transport in hadron-hadron collisions~\cite{Kharzeev:1996sq}, as discussed later. 

The photon-induced processes used in this analysis are inclusive $\gamma$+Au interactions selected from Au+Au collisions at the center-of-mass energy per nucleon pair (\sqrtsNN) of 54.4 GeV. In such collisions, quasi-real photons ($\gamma$), $i.e.$, those having almost zero virtuality ($q^2\sim0$), with  an average energy of $E_{\gamma}\simeq 0.8$ GeV are emitted from the projectile Au ion and interact with the target Au ion possessing an energy of 27.2 GeV per nucleon~\cite{vonWeizsacker:1934nji,Williams:1934ad,Wang:2022ihj}. This results in an average center-of-mass energy of $W_{\gamma N} \sim \SI{9}{GeV}$ for the photon-nucleon system~\cite{Lewis:2022arg}. The $\gamma$+Au events are selected by exploiting their characteristic asymmetric rapidity distribution of particle production and the differing neutron distributions of the target and projectile nuclear remnants \cite{SupplementalMaterial}.

Yields of protons and antiprotons as a function of $\mT-m_p$, where $m_{p}$ is the proton mass and $m_T=\sqrt{\pT^2+m_p^2}$ with $\pT$ denoting the transverse momentum, are measured in twelve different rapidity intervals\cite{SupplementalMaterial}. The net-proton yield is calculated as a proxy to quantify the baryon transport. In order to extract the total yields, proton, antiproton and net-proton spectra are parameterized and extrapolated to unmeasured regions\cite{SupplementalMaterial}.
\begin{figure}[ht]
\centerline{\includegraphics[width=0.8\linewidth]{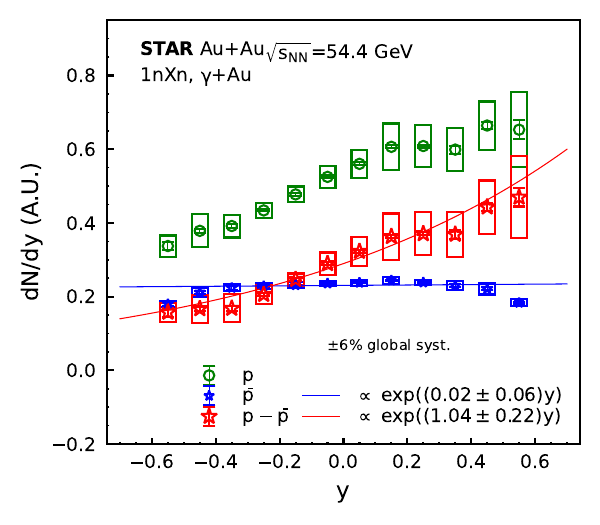}}
    \caption{{\bf Baryon transport in $\gamma$+Au interactions.} Proton, antiproton, and net-proton rapidity densities ($dN/dy$) in $\gamma$+Au events selected from \AuAu\ collisions at \sqrtsNN\ = 54.4 GeV. The selection requires remnants of a single neutron ($1n$) in the $\gamma$-going direction (negative $y$) and more than one neutron ($Xn$) in the Au-going direction (positive $y$). Fit results with an exponential function to antiproton and net-proton distributions are shown as solid curves. The 6\% global uncertainty is a correlated systematic uncertainty applied to all data points.}
    \label{fig:Photonuclear_dNdy}
\end{figure}
Figure~\ref{fig:Photonuclear_dNdy} shows the proton, antiproton and net-proton rapidity densities ($dN/dy$) in $\gamma$+Au events within $|y| < 0.6$ in the center-of-mass frame of the target and projectile Au nuclei. The efficiency for selecting $\gamma$+Au events is not accounted for. This missing normalization factor does not affect the extracted rapidity slopes but our reported absolute yields cannot be directly compared to theoretical calculations. In this figure, results from events where the photon travels in the negative $y$ direction are plotted unchanged, whereas those from the reversed collision kinematics
 are reflected around $y=0$ and added. This is because $\gamma$+Au events are selected from Au+Au collisions and either Au nucleus can emit the photon. 

\begin{figure}[ht!]
\centering
    \includegraphics[width=0.95\linewidth]{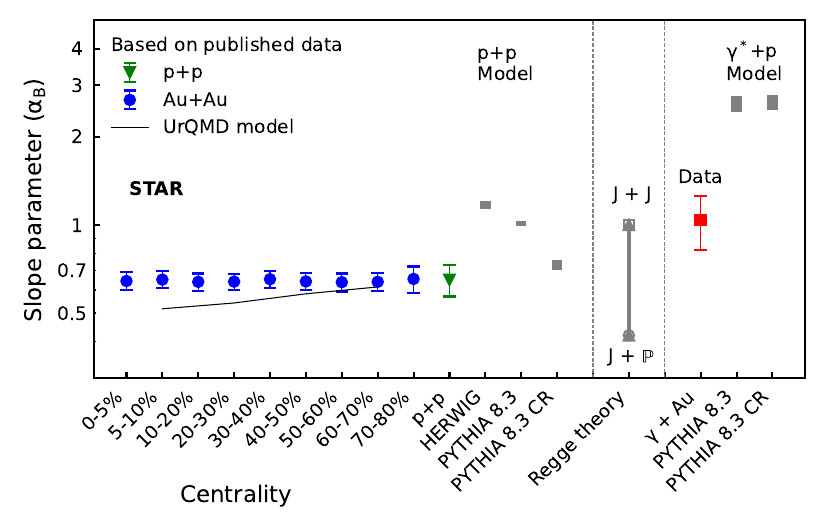}
    \caption{{\bf Rapidity slope of baryon transport.} The exponential slopes of the net-proton $dN/dy$ vs. $\Delta y$ in different centrality intervals of Au+Au collisions~\cite{Lewis:2022arg,STAR:2008med,STAR:2017sal} and $p$+$p$ collisions~\cite{Tsang:2024zsq}. They are compared to those from $\gamma$+Au events and predictions from models described in text. Vertical bars around experimental data points represent combined statistical and systematic uncertainties. The two points from Regge theory are connected by a gray band to show the range of $\alpha_B$ values, with the lower bound given by $J+\mathbf{P}$ and upper bound by $J + J$.}
    \label{fig:Photonuclear_dNdyGlobal}
\end{figure}

Both the proton and net-proton $dN/dy$ have positive slopes with respect to $y$. The antiproton rapidity distribution is mostly flat against $y$ with an exponential slope close to zero, {\it i.e.}, $\alpha_{\bar{B}}=0.02 \pm 0.05$. This is consistent with theoretical expectations of a weaker rapidity dependence for proton-antiproton pair production, the sole production mechanism for antiprotons, than for baryon transport~\cite{Artru:1990wq,Kharzeev:1996sq}, and is a clear demonstration that the strong rapidity dependence of the net-proton yield is not caused by the $\gamma$+Au event selection. A fit to the net-proton rapidity density with Eq.~\ref{eq:ydep} and $Y_{\mathrm{beam}}=4.06$ gives $\alpha_{B}=1.04 \pm 0.22$. The observed value is compared to model predictions in Fig. \ref{fig:Photonuclear_dNdyGlobal}, which shows that $\alpha_{B}$ for $\gamma$+Au events is consistent with the Regge theory prediction incorporating the baryon junction. In contrast, predictions from PYTHIA 8.3 default and color reconnection (CR) Mode 2 tunes~\cite{Christiansen:2015yqa}, based on the valence quark picture, significantly overpredict the measured slope for $\gamma$+Au events. Both tunes simulate the $\gamma^*$+$p$ process, where $\gamma^*$ denotes a virtual photon, with kinematics constrained to be similar to those of the $\gamma$+Au events. Specifically, the colliding protons have a beam energy of 27 GeV and the virtual photon kinematics are limited to $q^2 < 0.01~(\mathrm{GeV}/c)^2$ and $E_\gamma < 2$ GeV. The PYTHIA 8.3 default tune produces baryons mainly through the so-called ``popcorn" mechanism~\cite{Lonnblad:2023kft}, whereas the CR tune differs from the default by allowing the dynamical formation of baryon junctions through color reconnection prior to hadronization, a partial imitation of the baryon junction mechanism. However, it does not include junctions in colliding nucleons or implement mechanisms for junction scatterings. 

We now shift our focus to baryon transport along rapidity in hadronic processes. An exponential dependence of the net-proton yield on $\Delta y$ is also observed in \AuAu\ collisions~\cite{Lewis:2022arg}. In this case of symmetric collisions, net-proton yields at mid-rapidity are measured over a large range of beam energies ($\sqrtsNN = 7.7 - 200$ GeV)~\cite{STAR:2008med,STAR:2017sal} to extract the exponential dependence on $\Delta y=Y_{\mathrm{beam}}-y$ by fitting with Eq.~\ref{eq:ydep}. In this study, the measurements are made at $y\sim0$, while $Y_{\mathrm{beam}}$ varies with beam energy. This procedure is followed in Ref.~\cite{Lewis:2022arg}, and the resulting $\alpha_B$ as a function of centrality is shown in Fig.~\ref{fig:Photonuclear_dNdyGlobal}. Here, centrality refers to the geometric overlap between the two colliding nuclei, with central (peripheral) events referring to those with small (large) distances between the centers of the colliding nuclei at the time of overlap.  The most central (peripheral) collisions are labeled as 0-5\% (70-80\%) in Fig.~\ref{fig:Photonuclear_dNdyGlobal}. $\alpha_B$ is independent of collision centrality with an average value of $0.64 \pm 0.05$~\cite{Lewis:2022arg}, inconsistent with the valence quark transport that is expected to get enhanced by increasing multiple scatterings with centrality (peripheral to central) and/or collision energy \cite{Itakura:2003jp}. This is corroborated by the consistent $\alpha_B$ value in $p$+$p$ collisions, which is estimated to be $0.65\pm0.08$ with data that span over a beam energy range of $\sqrt{s}=23.4-\SI{200}{GeV}$~\cite{Tsang:2024zsq} and included in Fig.~\ref{fig:Photonuclear_dNdyGlobal}. The $\alpha_B$ value in Au+Au collisions is also consistent with the slope parameter for $\gamma$+Au processes within 1.7$\sigma$. Please note that $\alpha_B$ could depend on collision kinematics as predicted in~\cite{Artru:1974zn,Artru:1990wq}, but our current measurements in Au+Au and $\gamma$+Au events are not sufficient to draw firm conclusions on such kinematics dependence. 

Figure~\ref{fig:Photonuclear_dNdyGlobal} also includes the predicted slopes from the Regge theory for the baryon junction~\cite{Kharzeev:1996sq}, which encompass the Au+Au data. Slope parameters from Ultra relativistic Quantum Molecular Dynamics (UrQMD)\cite{Bass:1998ca,Bleicher:1999xi} simulations of Au+Au collisions and from PYTHIA 8.3 and HERWIG 7.2 simulations of $p$+$p$ collisions~\protect\cite{Lewis:2022arg}, all of which assume valence quarks as the carriers of the baryon number, are extracted the same way as for the Au+Au data and shown for comparison. A much stronger centrality dependence of $\alpha_B$ is seen in UrQMD than that found in data. It decreases by about 0.1 from peripheral to central collisions in UrQMD, as expected for the valence quark transport. The Au+Au data are consistent with PYTHIA 8.3 CR tune, but not with either the PYTHIA 8.3 default tune or Herwig 7.2 which employs a quark-diquark model for baryon production~\cite{Bellm:2015jjp}. On the other hand, as mentioned, both tunes of PYTHIA 8.3 overshoot the measured slope for $\gamma$+Au events. A full implementation of the baryon junction mechanism \cite{Artru:1974zn,Rossi:1977cy,Kharzeev:1996sq} by including baryon junctions in the incoming protons and junction scatterings could potentially improve the agreement between data and modeling. 

\section*{Baryon transport vs. charge transport}\label{Sec:BaryonVsCharge}
One can provide further insights into baryon number carrier by comparing charge and baryon transport. Because valence quarks are known to carry electric charge~\cite{Whalley:2003, Riordan:1992hr}, in the scenario of valence quarks also carrying the baryon number, the fraction of baryon number transported from projectile and target nuclei to mid-rapidity ($B/(2A))$ is expected to be equal to that of charge number transport of $Q/(2Z)$, where $Z$ and $A$ are the atomic and mass numbers of the colliding nuclei. That is $B/A\simeq Q/Z$. 
For the alternative scenario of the baryon junction as the carrier, one expects $B/A>Q/Z$ because junction carries zero electric charge and one unit of baryon number, and can be transported by strong interactions more efficiently. 

To experimentally quantify the charge and baryon number transport to a specific rapidity in high-energy nucleus-nucleus collisions, we measure net-charge ($Q$) and net-baryon ($B$) numbers at that rapidity. They are defined as:
\begin{linenomath*}
\begin{equation}
Q=(N_{\pi^{+}}+N_{K^{+}}+N_{p})-(N_{\pi^{-}}+N_{K^{-}}+N_{\bar{p}})
\label{eq:net-Q}
\end{equation}
and
\begin{equation}
B=(N_{p}+N_{n})-(N_{\bar{p}}+N_{\bar{n}}). 
\end{equation}
\end{linenomath*}
Here, $N_{\pi^{+}}$, $N_{K^{+}}$, $N_{p}$ are the yields of positively charged pions ($\pi$), kaons ($K$) and protons ($p$), and $N_{\pi^{-}}$, $N_{K^{-}}$, and $N_{\bar{p}}$ denote their antiparticles \protect\cite{SupplementalMaterial}. $N_{n}$ and $N_{\bar{n}}$ are the yields of neutrons and antineutrons.

As illustrated by Eq.~\ref{eq:net-Q}, the small net-charge number is determined from the difference between large yields of positively and negatively charged particles. Therefore, measuring charge transport in one species of nuclear collisions usually involves large uncertainties, making it difficult to achieve the necessary experimental precision. This challenge can be mitigated by evaluating the difference in the charge transport ($\Delta Q$) between two isobar collision species (denoted ``Iso1" and ``Iso2" below), $i.e.$, nuclei with same $A$ but different $Z$, via double ratios: 
\begin{linenomath*}
\begin{equation}
\Delta Q=Q_{\rm{Iso1}}-Q_{\rm{Iso2}}\approx N_{\pi}(R2_{\pi}-1)+N_{K}(R2_{K}-1)+N_{p}(R2_{p}-1),
\end{equation}
\end{linenomath*}
 where $N_{\pi}$, $N_{K}$, and $N_{p}$ are the average yields over positively and negatively charged hadrons and over the two collision systems. $R2_{\pi}=(N_{\pi^{+}}/N_{\pi^{-}})_{\rm{Iso1}}/(N_{\pi^{+}}/N_{\pi^{-}})_{\rm{Iso2}}$ is the double ratio for pions, and $R2_{K}$ and $R2_{p}$ are defined similarly for kaons and protons~\cite{Lewis:2022arg}. Such a relation between $\Delta Q$ and double ratios holds for isobar collision systems as they are expected to produce similar particle multiplicities ~\cite{Lewis:2022arg}. These double ratios, which deviate from unity on the order of $10^{-3}$~\cite{SupplementalMaterial}, need to be measured with negligible uncertainties to allow for a meaningful determination of the charge transport. Then the ratio $\langle B\rangle/A$ is compared against $\Delta Q/\Delta Z$, where $\langle B\rangle$ is the average net-baryon number in the two isobar collision systems, and $\Delta Z$ is the difference in $Z$ between the isobar nuclei. As a reminder, one expects $\langle B\rangle/A\times\Delta Z/\Delta Q=\langle B\rangle/\Delta Q\times\Delta Z/A\approx 1$ in the valence quark picture, and $\langle B\rangle/\Delta Q\times\Delta Z/A> 1$ in the junction picture.
 
 In 2018, the STAR experiment \cite{Ackermann:2002ad} recorded about 2 billion events each for two isobar collision species ( ${}^{96}_{44}\rm{Ru}$+${}^{96}_{44}\rm{Ru}$ and ${}^{96}_{40}\rm{Zr}$+${}^{96}_{40}\rm{Zr}$ ). This provides a unique opportunity to measure the charge transport with high precision, because these data were taken with almost identical accelerator and detector conditions \cite{STAR:2021mii} and can be analyzed with the same procedure, resulting in negligible systematic uncertainties in double ratios. For this pair of isobar nuclei, $\Delta Z=4$ and $A=96$.

Yields of positively and negatively charged $\pi$, $K$, $p$, as well as the double ratios ($R2_{\pi}$, $R2_{K}$, and $R2_{p}$) are measured as a function of \pT\ within $|y|<0.5$ in five centrality classes \cite{SupplementalMaterial}. These yields consist of primordial production from the collision zone as well as secondary products from resonance and weak decays. Neutrons cannot be measured by the STAR detector, therefore, their yields have to be estimated. The primordial production is determined based on the measured proton yields and antideuteron-to-deuteron ratios ($\bar{d}/d$), whereas the decay or feed-down contribution, contributing to approximately 30\% of the total neutron yield, is estimated from hyperon yields and known branching ratios~\protect\cite{SupplementalMaterial}. The same procedure is applied to estimate the yields for antineutrons~\protect\cite{SupplementalMaterial}. The yield distributions and double ratios are fit with a Blast-wave function \cite{Schnedermann:1993ws} and a linear function, respectively, to extrapolate from the measured to the full \pT\ range. $\langle B\rangle$ and $\Delta Q$ are calculated in individual \pT\ bins using the measured data or extrapolated values, and then integrated to obtain the final reported values. Because the used particle yields include decay contributions, the above calculated $\langle B\rangle$ and $\Delta Q$ values closely approximate the true net-baryon and net-charge numbers with a relative error of 1\% or less owing to unmeasured nuclei.

\begin{figure}[ht!]
    \centerline{\includegraphics{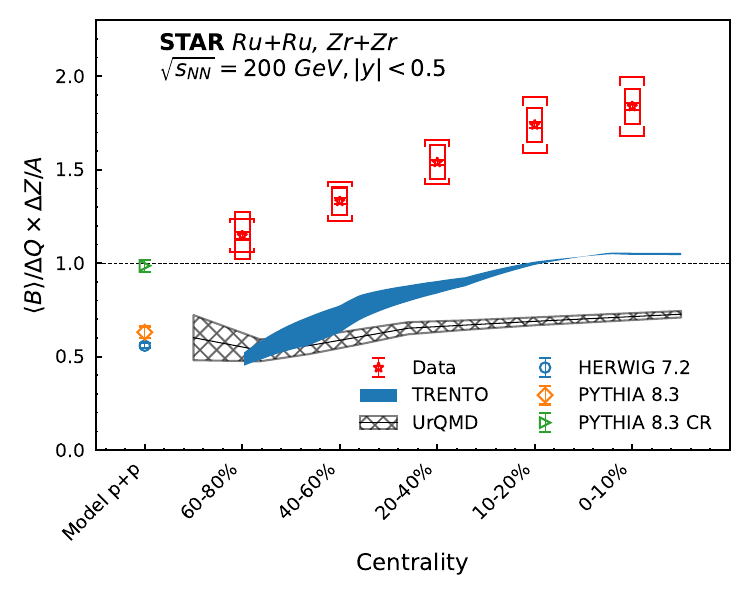}}
    \caption{{\bf Baryon to charge transport ratio.} The ratio of the mean net-baryon number ($\langle B\rangle$) to the net-charge difference between \RuRu\ and \ZrZr\ collisions ($\Delta Q$) at mid-rapidity ($|y| < 0.5$), scaled by $\Delta Z/A=4/96$, as a function of centrality. 
    Red stars denote data points; surrounding boxes and vertical bars indicate systematic and statistical uncertainties, respectively. Uncertainties arising from estimations of feed-down contributions to neutrons and antineutrons are shown separately as brackets. The horizontal dashed line at 1 shows the naive expectation if the baryon number is carried by valence quarks and all other effects are ignored. Model calculations from UrQMD (hatched band)\protect\cite{Bass:1998ca,Bleicher:1999xi,Lv:2023fxk}, TRENTO (solid band)~\protect\cite{Xu:2021qjw}, PYTHIA 8.3 (diamond)~\protect\cite{Christiansen:2015yqa}, PYTHIA 8.3 with CR Mode 2 (triangle), and HERWIG 7.2 (circle)~\protect\cite{Bellm:2015jjp} are also shown for comparison. See text for details.}
    \label{fig:Isobar}
\end{figure}

Figure \ref{fig:Isobar} shows the resulting $\langle B\rangle/\Delta Q$, multiplied by $\Delta Z/A$, as a function of centrality at mid-rapidity ($|y|<0.5$), corresponding to an average rapidity shift of 5.36. In 0-10\% most central collisions, $\langle B\rangle/\Delta Q\times\Delta Z/A= 1.84$ $\pm\ 0.02$ (stat.) $\pm\ 0.09$ (syst.) $\pm\ 0.16$ (feed-down), much larger than the naive expectation of unity for the scenario of valence quarks carrying the baryon number. Larger baryon transport compared to charge transport is consistent with the expectation of the baryon junction scenario. The ratio decreases monotonically from 0-10\% most central to 70-80\% peripheral collisions. A similar decreasing trend is seen in the TRENTO model calculation \cite{Xu:2021qjw}, which only geometrically accounts for the protons and neutrons participating in the collision without implementing any baryon transport mechanism. For this model specifically, $\langle B\rangle$ is taken as the sum of proton and neutron numbers, and $Q$ is taken as the proton number. The decreasing trend is attributed to the larger neutron skin for the Zr nucleus than that for Ru, which results in an increasingly smaller fraction of participating protons in \ZrZr\ collisions compared to \RuRu\ as the collisions become more peripheral. The width of the band for the TRENTO calculation corresponds to the uncertainty in the neutron skin values used. More sophisticated calculations based on UrQMD~\cite{Bass:1998ca,Bleicher:1999xi}, which simulates the entire evolution of nuclear collisions and can reproduce the net-baryon yields in central Ru+Ru and Zr+Zr collisions \cite{SupplementalMaterial}, are also shown for comparison. It predicts a value of 0.5 to 0.7, significantly below the experimental data. 
This ratio is smaller than the naive expectation of unity, owing to an excess of antistrange quarks over strange quarks at mid-rapidity in the simulation~\cite{Ross:2025qxr}. The neutron skin effect is not included in the UrQMD calculation, hence it does not exhibit the significant increasing trend from peripheral to central collisions discussed above. The limiting case of $p$+$p$ collisions at $\sqrt{s}=$ 200 GeV is simulated with three event generators: HERWIG 7.2~\cite{Bellm:2015jjp}, PYTHIA 8.3 (default tune)~\cite{Bierlich:2022pfr}, and PYTHIA 8.3 with CR Mode 2~\cite{Bierlich:2022pfr,Christiansen:2015yqa}, for which $B/Q$ ratios instead of $B/\Delta Q$ are shown.
Both HERWIG 7.2 and PYTHIA 8.3 default tune predict $B/Q$ values between 0.5-0.6, whereas PYTHIA 8.3 CR enhances baryon production through its partial imitation of baryon junction, resulting in $B/Q=0.99\pm0.03$. Although the number of participating nucleons for peripheral Ru+Ru and Zr+Zr collisions is close to that of $p$+$p$ collisions, the peripheral data are affected by the aforementioned neutron skin effect and should not be directly compared to HERWIG and PYTHIA predictions. 

\section*{Discussions and outlook} 

We have reported new measurements from $\gamma$+Au and isobar collisions with the STAR detector that investigate whether the carrier of the baryon number is valence quarks or alternative topological entities such as the baryon junction. These results, supported by data previously reported by STAR from Au+Au collisions at a variety of beam energies, disfavor the valence quark picture. Further investigations into existing and alternative theories are warranted; to be viable, such theories must simultaneously explain all observed phenomena, and so far, only the baryon junction framework remains qualitatively consistent~\cite{Pihan:2024lxw}. Future measurements at the existing Continuous Electron Beam Accelerator Facility, Large Hadron Collider and future Electron-Ion Collider will further shed light on the nature of the baryon number carrier and directly probe its distribution function in nuclei~\cite{Lewis:2022arg,EICAccardi:2012qut,Cebra:2022avc,Frenklakh:2023pwy}.


\renewcommand{\refname}{References and Notes}
\bibliographystyle{naturemag}
\bibliography{bibliography}

\section*{Acknowledgments}

We thank Professors Dmitri Kharzeev and Ziwei Lin for their valuable discussions. We thank Center for Frontiers in Nuclear Science for hosting a dedicated workshop on baryon dynamics in January 2024. We thank the RHIC Operations Group and SDCF at BNL, the NERSC Center at LBNL, and the Open Science Grid consortium for providing resources and support.  \textbf{Funding}: This work was supported in part by the Office of Nuclear Physics within the U.S. DOE Office of Science, the U.S. National Science Foundation, National Natural Science Foundation of China, Chinese Academy of Science, the Ministry of Science and Technology of China and the Chinese Ministry of Education, NSTC Taipei, the National Research Foundation of Korea, Czech Science Foundation and Ministry of Education, Youth and Sports of the Czech Republic, Hungarian National Research, Development and Innovation Office, New National Excellency Programme of the Hungarian Ministry of Human Capacities, Department of Atomic Energy and Department of Science and Technology of the Government of India, the National Science Centre and WUT ID-UB of Poland, German Bundesministerium f\"ur Bildung, Wissenschaft, Forschung and Technologie (BMBF), Helmholtz Association, Ministry of Education, Culture, Sports, Science, and Technology (MEXT), Japan Society for the Promotion of Science (JSPS), and Agencia Nacional de Investigacion y Desarrollo de Chile (ANID), Chile. \textbf{Author contributions}:  This paper is produced collectively by the STAR Collaboration. All individuals who have made significant contributions to the paper are listed alphabetically as authors, per the STAR’s Publication Policy. \textbf{Competing interests}: The authors declare that they have no competing interests. \textbf{Data and materials availability}: The codes used to process the raw data collected by the STAR detector are publicly available on GitHub (https://github.com/star-bnl). The codes used to process the reconstruction data into the final data points presented in this work are not publicly available but are archived at the Brookhaven National Laboratory computing facility. The data are deposited in HepData~\cite{hepdata.154708}. The analysis codes used to generate the theoretical curves and fit functions shown in the figures are deposited in Zenodo~\cite{zenodo}.

\clearpage
\input{Supplemental}

\end{document}

%% file: author.tex
\author{
B.~E.~Aboona$^{1}$,
J.~Adam$^{2}$,
L.~Adamczyk$^{3}$,
I.~Aggarwal$^{4}$,
M.~M.~Aggarwal$^{4}$,
Z.~Ahammed$^{5}$,
A.~K.~Alshammri$^{6}$,
E.~C.~Aschenauer$^{7}$,
S.~Aslam$^{8}$,
J.~Atchison$^{9}$
% ...
}

\address{\rm{(STAR Collaboration)}}

\address{$^{1}$Texas A\&M University, College Station, Texas 77843}
\address{$^{2}$Czech Technical University in Prague, FNSPE, Prague 115 19, Czech Republic}
\address{$^{3}$AGH University of Krakow, FPACS, Cracow 30-059, Poland}
\address{$^{4}$Panjab University, Chandigarh 160014, India}
\address{$^{5}$Variable Energy Cyclotron Centre, Kolkata 700064, India}
\address{$^{6}$Kent State University, Kent, Ohio 44242}
\address{$^{7}$Brookhaven National Laboratory, Upton, New York 11973}
\address{$^{8}$Fudan University, Shanghai, 200433 }
\address{$^{9}$Abilene Christian University, Abilene, Texas   79699}
% ...

%% file: Supplemental.tex











\setcounter{figure}{0}
\setcounter{equation}{0} 
\renewcommand{\theequation}{S\arabic{equation}} 
\renewcommand{\thefigure}{S\arabic{figure}}

\thispagestyle{empty}

\begin{center}

{\LARGE Supplementary Materials for}

\vspace{0.6cm}

{\Large \textbf{Tracking the baryon number with nuclear collisions}}

\vspace{0.8cm}

{\large STAR Collaboration}

\vspace{0.6cm}

{\large Corresponding author: \textcolor{blue}{star-publication@bnl.gov}}

\end{center}

\vspace{2.5cm}

\noindent{\large \textbf{The PDF file includes:}}

\vspace{0.5cm}

\begin{itemize}
\item Materials and Methods
\item Figs. S1 to S5
\item References (49-64)
\end{itemize}

\clearpage

\section{Materials and Methods}

\subsection{Baryon transport in photonuclear events}\label{Sec:GammaASupplemental}
Studying baryon transport in $\gamma$+Au interactions provides another avenue for identifying the baryon number carrier.

The key to selecting $\gamma$+Au events is the asymmetric rapidity dependence of produced particles. Model calculations indicate that in such collisions a large amount of activity is expected in the Au-going direction due to nucleus fragmentation, while little-to-no activity is expected in the $\gamma$-going direction~\cite{Lewis:2022arg}. Most importantly, a few neutrons from the fragmenting Au nucleus are expected to be traveling in the Au-going direction, while the $\gamma$-going side is expected to see one or two neutrons due to Coulomb excitations. In addition, the rapidity distribution of charged tracks around mid-rapidity is expected to be strongly asymmetric, with more particles in the Au-going direction. 

Following those guidelines, to trigger on inclusive $\gamma+$Au interactions from Au+Au collisions, we measure the energy distribution of neutrons detected by the Zero Degree Calorimeter detectors (ZDCs). We then identify peaks in the energy distribution that correspond to different numbers of neutrons ($1n$, $2n$, $\dots$, $Xn$). We require the ZDC detector in the $\gamma$-going side to have a single neutron hit ($1n$) and more than one neutron ($Xn$ where $(X\!>\!1)$) in the Au-going direction. Requiring a single neutron hit ($1n$) instead of zero neutrons ($0n$) in the $\gamma$-going direction helps reduce backgrounds from the beam hitting the beam pipe and beam-gas events, which can also produce rapidity asymmetry of charged particles around mid-rapidity but their fragmenting neutrons would not reach the ZDC. 
Additional asymmetric cuts on the energy deposition in the Beam-Beam Counter (BBCs) also help to improve the purity of the $\gamma$+Au interactions by selecting events with low activity in the $\gamma$-going direction and high activity in the Au-going direction. Contamination due to hadronic events can be further reduced by exploiting the timing information from the Vertex Position Detectors (VPDs)~\cite{Llope:2014nva}. Unlike \AuAu\ collisions, in $\gamma$+Au interactions one of the VPDs sees little activity and so the timing difference between the two VPDs cannot be used to accurately determine the vertex position along the beam direction ($z$). We exploit this characteristic and require significant disagreement between the vertex $z$ positions measured by the Time Projection Chamber (TPC)~\cite{Anderson:2003ur} and VPDs: $\Delta V_{z}=|V_{z}^{\rm{TPC}} - V_{z}^{\rm{VPD}}|> 10$ cm. The TPC is a gaseous detector, which tracks the trajectories of traversing charged particles. For a more detailed description of this trigger, see Ref.~\cite{Lewis:2022arg}.

With these cuts we select about 2 million $\gamma$+Au events from about 100 million minimum bias Au+Au events at \sqrtsNN\ = 54.4 GeV. Due to the selection criteria used, this data sample is biased toward high multiplicity $\gamma$+Au interactions, where the average number of charged tracks within $|y|<0.5$ is about 7.


Protons and antiprotons are measured as TPC tracks required to have at least 15 TPC space points used for track reconstruction, and each track has a Distance of Closest Approach (DCA) to the collision vertex less than 3 cm. 
The proton and antiproton yields are extracted by using the $m^{2}$ distribution of charged tracks measured by Time-Of-Flight (TOF) detector \cite{Llope:2003ti}, and requiring $0.7 < m^{2} < 1.06~\mathrm{GeV}^2/c^4$. 
Based on the arrival time ($t_{f}$) measured in the TOF, the collision start time ($t_{0}$) measured in the VPDs, the momentum ($p$) from the TPC, and the trajectory length ($L$) from collision vertex to TOF, one can calculate the particle's mass squared: $m^{2} = p^{2}[c^{2}(t_f-t_0)^{2}/L^{2}-1]$, where $c$ is the speed of light.
They are corrected for the $m^{2}$ cut efficiency, evaluated by fitting the $m^2$ distribution with a Student’s t-function. As aforementioned, VPDs supply collision start time for the $m^{2}$ calculation, and the VPD signal from only the Au-going side can still provide sufficient timing resolution for an accurate start time determination. The presented results focus on total yields of protons and antiprotons, and therefore contributions from weak decays are not subtracted.

Measured proton and antiproton yields are corrected for various contaminations. The first correction is related to the contamination from peripheral hadronic Au+Au events, dominantly those in the 80-100\%  centrality class. The fraction ($r$) of peripheral events in the selected $\gamma$+Au sample is estimated to be about 11-13\% depending on which colliding Au nucleus is chosen as the photon emitter. Due to the complications in correcting for the trigger bias, it is difficult to measure particle yields in collisions that are more peripheral than 80\%, so protons and antiprotons from 60-80\% peripheral collisions are used as an estimate of the peripheral background ($Y_{p}^{AA}$). A multiplicative factor ($f$) is used to account for the multiplicity difference between 60-80\% and 80-100\% events. The corrected yield for $\gamma+$Au processes ($Y_{p}^{\gamma A }$) is as follows:

\begin{equation}
    Y_{p}^{\gamma A }     = \frac{Y_{p}^{\gamma A +AA} - rf \, Y_{p}^{AA}}{1-r},
\end{equation}
where $Y_{p}^{\gamma A +AA}$ denotes the proton or antiproton yield in the selected $\gamma+$Au events.

\begin{figure}[h]
\centering
    \includegraphics[width=1.0\linewidth]{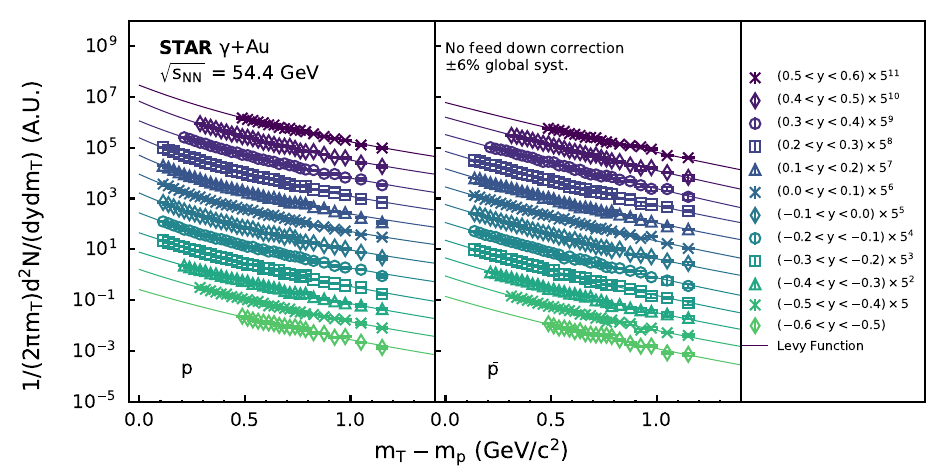}
    \caption{{\bf (Anti)proton spectra in $\gamma$+Au.} The proton and antiproton $m_T$ spectra in $\gamma+$Au processes selected from \AuAu\ collisions at \sqrtsNN\ = 54.4 GeV. Solid curves represent fits to the spectra with the L\'evy function. There is an additional 6\% global uncertainty not included in the figure. See text for details.  }
    \label{fig:Supplemental_GammaASpectra}
\end{figure}

Corrections for the tracking efficiency and energy loss in the TPC are evaluated using the embedding technique, where simulated particles of interest are propagated through the GEANT 3 \cite{Brun:1987ma} simulation of the STAR detector, mixed with real events and reconstructed the same way as data. The TOF matching efficiency correction is estimated with a data-driven method, and applied to account for the efficiency of the TOF detector in detecting tracks previously measured by the TPC. Additional protons can be knocked out of the detector material by high-\pT\ hadrons, and thus need to be subtracted. Their contribution is estimated by fitting the DCA distribution for the measured proton sample with two components: one for the signal protons using the DCA distribution for antiprotons which do not suffer from knock-out contamination, and the other for the background protons parameterized with a functional shape. 
The proton and antiproton spectra for $\gamma+$Au collisions, shown in Fig.~\ref{fig:Supplemental_GammaASpectra}, are the average spectra for the cases when the photon is emitted by either of the two colliding Au nuclei. We define the proton rapidity under the condition that the photon always travels in the negative $y$ direction.  The solid curves represent L\'evy function fits used to extract the rapidity density ($dN/dy$). The L\'evy function is a generalized exponential function: $A \bigl( 1 + \frac{\mT- m_p}{Cn}\bigr)^{-n}$, where $m_{p}$ is the proton mass, $m_T=\sqrt{\pT^2+m_p^2}$ is the transverse mass, $A$ is a normalization factor, $C$ is a characteristic transverse mass scale, and $n$ is a shape parameter that controls the slope of the spectra at high \mT.

The total relative systematic uncertainties of the $dN/dy$ values are on average about 13\% for protons and about 8\% for antiprotons. It is dominated by the uncertainties due to the $\gamma$+Au event selection. We find the difference in the $dN/dy$ values between events where either Au nucleus in Au+Au collisions emits the photon (4\% for protons and negligible for antiprotons). We also change the selection on the VPD, requiring no activity on the $\gamma$-going side instead of the default cut of requiring significant disagreement between the vertex $z$ positions as measured by the TPC and VPD detectors (8\% for protons and 3\% for antiprotons). Since requiring no VPD signal corresponds to an extreme case given detector noises, differences to the default results are divided by $\sqrt{3}$, which are then used as uncertainty estimates. The uncertainty due to  extrapolation is estimated by comparing the $dN/dy$ values extracted with a Blast-wave function to the default results and contributes about 5\% for protons and 2\% for antiprotons. Various other sources of systematic uncertainties are also considered. These sources include tracking efficiency uncertainty (5\%), contribution of beam gas and beam pipe events (3\%), track selection uncertainty (1\%) (using different criteria), peripheral event contamination (4\%), knock-out proton correction uncertainty ($<$1\% for protons), uncertainty in track background from $m^2$ fits ($<0.1\%$), uncertainty related to particle identification (2\% for protons and 5\% for antiprotons), the uncertainty of the efficiency correction for the $m^2$ cut ($<$1\%) and missing weak decays (0.6\%). Among them, uncertainties in tracking efficiency and contribution of beam gas and beam pipe events apply to all data points, and thus classified as the global uncertainty amounting to 6\%. 

When fitting the net-proton $dN/dy$ distribution to extract the slope parameter, we separate the systematic uncertainties into those which affect the $dN/dy$ distribution with a clear $y$ dependence (``correlated") and those that do not (``uncorrelated"). To find the central value of the slope, we fit the $dN/dy$ values with an exponential function using the uncorrelated systematic uncertainties and statistical uncertainties added in quadrature, accounting for approximately 40\% of the total uncertainty. The correlated systematic uncertainty sources include the peripheral contamination and the photonuclear event selection. We perform separate fits on the $dN/dy$ values which have been shifted by each of the correlated systematic uncertainties. 
 The resulting differences between the default value and shifted slopes are taken as additional uncertainties.


\subsection{Charge and baryon transport in \RuRu\ and \ZrZr\ collisions}
\label{sect:SM-isobar-spectra}
\RuRu\ and \ZrZr\ collisions recorded by the STAR experiment~\cite{STAR:2021mii,YangLi_2023} are utilized for comparing charge and baryon transport.

\subsubsection{Invariant yields of charged particles}
\RuRu\ and \ZrZr\ events are retained if their collision vertices fall within $-35$ to $+25$ cm along the beam axis and within 2 cm radially from the STAR detector center.
Invariant yields of $\pi$, $K$, $p$, and $d$ as a function of transverse momentum (\pT) in different centrality classes are measured utilizing the TPC and the TOF. The event centrality is classified by fitting the distribution of charged particle multiplicity within the pseudorapidity range of $|\eta| < 0.5$ with the Monte Carlo Glauber model~\cite{STAR:2021mii}. Central (peripheral) events correspond to those with large (small) multiplicities and large (small) number of participating nucleons (\npart). 
Charged particle tracks are selected using criteria as detailed in Sec.~\ref{Sec:GammaASupplemental}.  
Furthermore, the ratio of the number of used TPC space points to the maximum possible number along the track trajectory is required to be at least 0.52 to remove split tracks. The kinematic coverage of the measurement is $0.2 <\pT < 2.5$ \gev\ for $\pi$ and $K$, $0.45 < \pT < 2.5$ \gev\ for $p$, and $0.9 < \pT < 4.2$ \gev\ for $d$, within the rapidity range of $|y| < 0.5$ for all species. 

Charged particles lose energy in the TPC by ionizing gas atoms. This energy loss ($dE/dx$), which depends on particle mass, enables particle identification. The variable $n\sigma_{i}=(\ln(dE/dx)_{\mathrm{measured}}-\ln(dE/dx)_{i, \mathrm{expected}})/R(\ln(dE/dx))$ is used, where $(dE/dx)_{\mathrm{measured}}$ is the measured energy loss in the TPC, $(dE/dx)_{i, \mathrm{expected}}$ is the expected energy loss for particle $i$ based on the Bichsel formalism \cite{Bichsel:2006cs}, and $R(\ln(dE/dx))$ is the experimental resolution of $\ln(dE/dx)$. To ensure good resolution, each track should have at least 10 TPC space points used for calculating $dE/dx$. Since the particle identification capability of the TPC is limited to low \pT, it is used up to 0.5 \gev, 0.25 \gev, and 0.7 \gev\ for $\pi$, $K$, $p$, respectively, above which the TOF is used. For the $d$ measurement, the TOF is always used. The raw yield for particle $i$ in a given \pT\ bin is extracted by counting the entries under the peak of particle $i$ in the $n\sigma_{i}$ distribution within $[\mu_{i}-2\sigma_{i}, \mu_{i}+2\sigma_{i}]$. $\mu_{i}$ and $\sigma_{i}$ are the mean and width of the signal peak, determined by fitting the $n\sigma_{i}$ distribution with a multi-Gaussian function. The missing yield outside the counting range is corrected for based on the fitted Gaussian function shape. To minimize the uncertainties in measuring the double ratios, the $n\sigma_{i}$ distribution for negatively charged particles in \RuRu\ collisions is fit first, and the resulting mean and width values for the various Gaussian distributions are used to fix the corresponding parameters when fitting positively charged particles in \RuRu\ collisions and particles of both charges in \ZrZr\ collisions.

The TOF is used for identifying $\pi$, $K$, $p$, at higher \pT\ ranges. 
A fit to the $m^{2}$ distribution in each \pT\ bin is performed using three Student's $t$-distributions, representing the three particle species, and the raw yields are obtained directly from the fit. Since the deuteron peak in the $m^{2}$ distribution is far away from other particles, a single student's $t$-distribution describing the signal plus an exponential function representing the background are used for fitting. The $d$ yields are also taken from the fit. As for the case of $n\sigma_{i}$ fitting, the shape parameters for negative particles in \RuRu\ collisions are used to fix the functional shapes for other cases while the yields are left as free parameters.

Corrections for the TPC tracking efficiency and energy loss, TOF matching efficiency and knock-out proton contamination are evaluated similarly as outlined in Sec.~\ref{Sec:GammaASupplemental}.  
A similar procedure is applied to subtract knock-out deuterons from the measured sample. For the (anti-)$d$ measurement, an additional absorption correction is applied to account for the overestimation of the (anti-)$d$ tracking efficiency due to GEANT 3 lacking the correct (anti-)$d$ interaction cross section in detector material~\cite{STAR:2019sjh, STAR:2001pbk}.

\begin{figure}
\centering
    \includegraphics[width=1.0\linewidth]{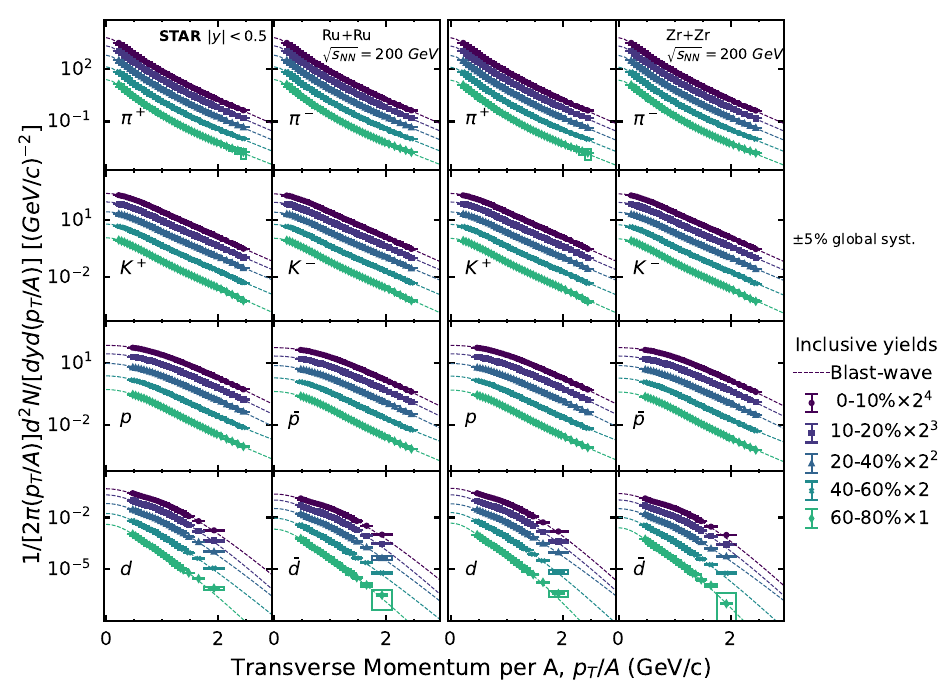}
\caption{{\bf Identified particle spectra.} Identified particle spectra, including all decay contributions, within $|y| < 0.5$ in \RuRu\ (left two columns) and \ZrZr\ (right two columns) collisions at \sqrtsNN\ = 200 GeV. From top to the bottom, the particles being plotted are pions, kaons, protons and deuterons, respectively. The $x$-axis, \pT$/A$, is the transverse momentum per nucleon. For simplicity, $A=1$ for pions and kaons even though they are not nucleons. For both the Zr and Ru cases, the left column shows positively charged particles, and the right shows negatively charged particles. The vertical error bar size represents the statistical uncertainty while the box height represents the systematic uncertainty. The dashed curves show the fitted Blast-wave function \protect\cite{Schnedermann:1993ws}. }
\label{fig:RuZrSpectrum}
\end{figure}

Fully corrected invariant yields of $\pi$, $K$, $p$, and $d$ within $|y| < 0.5$ are shown in Fig.~\ref{fig:RuZrSpectrum} as a function of \pT\ (\pT/2 for $d$) for \RuRu\ and \ZrZr\ collisions at \sqrtsNN\ = 200 GeV, respectively. Decay daughters of resonances and weak-decaying particles are included in the measured yields. Spectra in different centrality classes are scaled vertically with arbitrary factors for visual clarity. Vertical bars and boxes around data points, which are smaller than the marker size in most cases, represent statistical and systematic uncertainties. Systematic uncertainties on the yields include those originating from signal extraction, missing yields from weak decays due to the finite DCA cut of 3 cm, imperfect simulation of the detector response and, for the case of the proton measurement, the estimated knock-out background in the detector material. Among them, the 5\% uncertainty in the TPC track reconstruction efficiency, part of the detector simulation uncertainty, is not shown as this source applies to all particle species and is fully correlated across different \pT\ and centrality bins. 
Dashed curves crossing data points indicate fits to the measurements with a Blast-wave function \cite{Schnedermann:1993ws}. 

\begin{figure}[h]
\centering
    \includegraphics[width=1.0\linewidth]{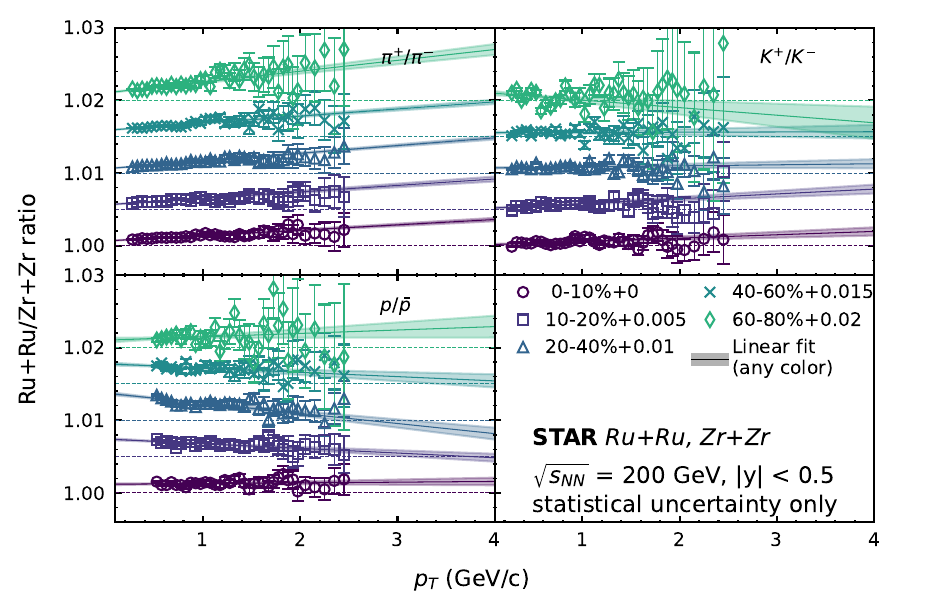}
    \caption{{\bf Particle yield double ratios.} Double ratios of $\pi$, $K$, and $p$ between positively and negatively charged particles and between \RuRu\ and \ZrZr\ collisions at \sqrtsNN\ = 200 GeV. Horizontal dashed lines display offsets for each centrality class, while solid lines with shaded error bands represent fits to the double ratios with a first-order polynomial function. }
    \label{fig:DoubleRatio}
\end{figure}

Double ratios of $\pi$, $K$, and $p$ ($R2_{\pi}$, $R2_{K}$, $R2_{p}$) between positively and negatively charged particles and between \RuRu\ and \ZrZr\ collisions are shown in Fig. \ref{fig:DoubleRatio} as a function of \pT\ in five different centrality classes. Only statistical uncertainties are displayed as the systematic uncertainties are negligible due to the cancellation of detector effects. Solid curves illustrate fits to the double ratios with a first-order polynomial function for extrapolating to both lower and higher \pT. Bands around solid lines represent the uncertainties on the fits originating from the statistical errors of the double ratios. Additional uncertainties arising from the extrapolation of double ratios to low and high $p_T$ in calculating the charge number are evaluated by changing the fit function to either a zeroth-order or a second-order polynomial function. Furthermore, impact of knock-out protons on the proton double ratio measurement is also included as a source of uncertainties.   


\subsubsection{Derivation of neutron and antineutron yields}
\label{sect:SM-neutron}
Neutrons make up a large fraction of baryons at mid-rapidity. However, they are not directly measured by the STAR detector, so their yields need to be estimated. The neutron yield ($N_{n}^\text{inc}$) consists of primordial production ($N_{n}^\text{pri}$) and feed-down contribution from decays of hyperons  ($N_{n}^\text{FD}$), which are baryons with at least one strange valence quark, $i.e.$,
\begin{equation}
N_{n}^\text{inc}=N_{n}^\text{pri}+N_{n}^\text{FD}.
\end{equation}


The feed-down contribution to the neutron yield is estimated based on the yields of hyperons: 
\begin{equation}
N_{n}^\text{FD} = 0.361(N_{\Lambda} + N_{\Xi^0} + N_{\Xi^-} + N_{\Omega}) + 0.4831 N_{\Sigma^+} + N_{\Sigma^-},
\label{eq:hyperons}
\end{equation}
where the numeric factors represent branching ratios to neutrons~\cite{ParticleDataGroup:2022pth} and the subscript of $N$ shows which hyperon species the yield corresponds to. Since $\Sigma^0$ decays 100\% to $\Lambda$ electromagnetically and is indistinguishable from $\Lambda$ experimentally~\cite{STAR:2002wby}, its yield is considered part of the $\Lambda$ yield ($N_\Lambda$).
The hyperon yields are derived based on the experimental measurement of feed-down contribution to protons as well as model calculations. Feed-down contributions to the proton sample can be written as:
\begin{equation}
N_{p}^\text{FD} = 0.639(N_{\Lambda} + N_{\Xi^0} + N_{\Xi^-} + N_{\Omega}) + 0.5157 N_{\Sigma^+}.
\label{eq:pFD}
\end{equation}
The fraction of protons that comes from feed-down has been evaluated as a function of $\langle N_\text{part}\rangle$ for Au+Au collisions at \sqrtsNN\ = 200 GeV using a data-driven method~\cite{star:2006ib}, and is extrapolated to the $\langle N_\text{part}\rangle$ values of \RuRu\ and \ZrZr\ collisions used in this analysis. $N_{p}^\text{FD}$ is equal to the observed proton yields in \RuRu\ and \ZrZr\ collisions multiplied by the extrapolated feed-down fraction. The yield ratios of different hyperons to that of $\Lambda$ are calculated using the THERMUS thermal model~\cite{WHEATON2009}, whose parameters are fitted to the yields of pion, kaon, proton, their antiparticle counterparts as well as the antideuteron-to-deuteron ratios measured in \RuRu\ and \ZrZr\ collisions. With these, yields for all hyperons can be evaluated using Eq.~\ref{eq:pFD}, and $N_{n}^\text{FD}$ is then estimated by Eq.~\ref{eq:hyperons}.

In the framework of the statistical thermal model and under Boltzmann approximation, the primordial production yield for a particle is given by~\cite{Senger:2004xa}: 
\begin{equation}
N^\text{pri}=F(m)e^{B\mu_{B}+S\mu_{S}+Q\mu_{Q}},
\end{equation}
where $F(m)$ is a function of the particle mass ($m$). $B$, $S$, and $Q$ are the baryon number, strangeness, and electric charge of the particle, while $\mu_{B}$, $\mu_{S}$, and $\mu_{Q}$ are the chemical potentials of the corresponding conserved quantum numbers. Consequently, 
\begin{align}
N_{\bar{p}}^\text{pri}&=F(m_{p})e^{-\mu_{B}-\mu_{Q}},\\
N_{d}&=F(m_{d})e^{2\mu_{B}+\mu_{Q}},\\
N_{\bar{d}}&=F(m_{d})e^{-2\mu_{B}-\mu_{Q}},\\
N_{n}^\text{pri}&=F(m_{n}\approx m_{p})e^{\mu_{B}}=N_{\bar{p}}^\text{pri}\sqrt{N_{d}/N_{\bar{d}}}.
\label{eq:nddbar}
\end{align}
In this equation, the subscripts $d$ and $p$ represent deuteron and proton, respectively. Although we derive the equation using the thermal model, the relationship holds in general as long as the deuteron yield is proportional to the yield product of neutrons and protons, as in the coalescence model of nuclei~\cite{Schwarzschild:1963zz, Gutbrod:1976zzr}. To obtain the primordial antiproton yield, the feed-down fraction to antiproton is evaluated as $N_{\bar{p}}^\text{FD}/N_{\bar{p}}^\text{inc} = N_{p}^\text{FD}/N_{p}^\text{inc}\times(N_{\bar{\Lambda}}/N_\Lambda)/(N_{\bar{p}}^\text{inc}/N_p^\text{inc})$. To reach this conclusion, we approximate all antihyperon-to-hyperon ratios as $N_{\bar{\Lambda}}/N_\Lambda$. The $N_{\bar{p}}^\text{inc}/N_p^\text{inc}$ and $N_{\bar{\Lambda}}/N_\Lambda$ ratios are extracted from Refs.~\cite{STAR:2008med, star2007sp}. 

The antineutron yields are estimated similarly as for neutrons. For the primordial production, $N_{\bar{n}}^\text{pri}=N_{p}^\text{pri}\sqrt{N_{\bar{d}}/N_{d}}$. Uncertainties arising from those in feed-down fractions, hyperon ratios, and the possible imbalance of proton and antiproton feed-down fractions are accounted for in the final results.

\begin{figure}[htbp]

    \centering
    \includegraphics[width=\linewidth]{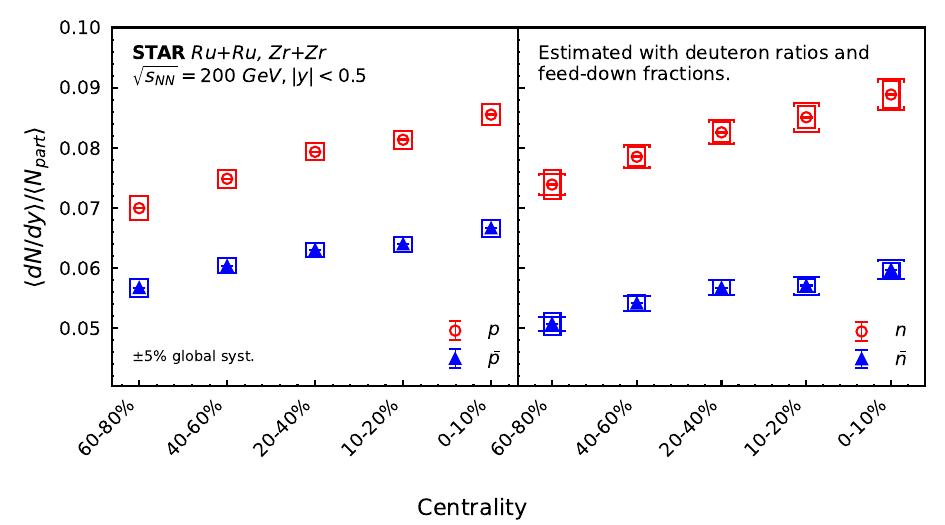}
    \caption{{\bf Proton and neutron yields.} Yields of protons (left) and neutrons (right), as well as their antiparticles, as a function of centrality. Proton and antiproton yields are measured within $0.45 < \pT < 2.5$ \gev\ and extrapolated to the full \pT\ region, while the neutron and antineutron yields are solely estimated. Brackets around data points in the right panel represent uncertainties in the estimation of feed-down contribution, while the boxes in both panels show systematic uncertainties from all other sources.}
    \label{fig:ProtonNeutronYield}
\end{figure}

The estimated neutron and antineutron yields, averaged over \RuRu\ and \ZrZr\ collisions and divided by \npart~\cite{STAR:2021mii} to take out trivial geometric effects, are plotted in the right panel of Fig.~\ref{fig:ProtonNeutronYield} as a function of centrality. Here, 0-10\% denotes the most central collisions while 60-80\% represents peripheral collisions. For comparison, yields for protons and antiprotons are also plotted in the left panel.

\subsubsection{Charge and baryon transport}
\begin{figure}[htbp]
\begin{minipage}{0.49\linewidth}
\centerline{\includegraphics[width=0.95\linewidth]{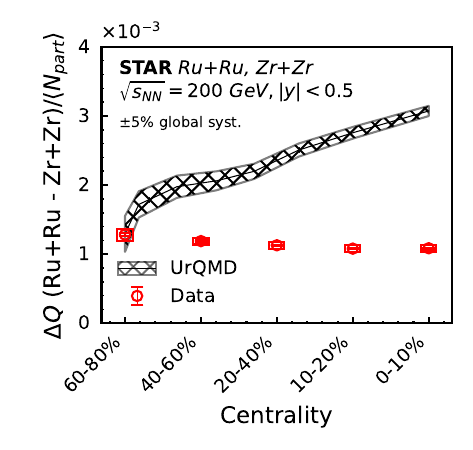}}
\end{minipage}
\begin{minipage}{0.49\linewidth}
\centerline{\includegraphics[width=0.95\linewidth]{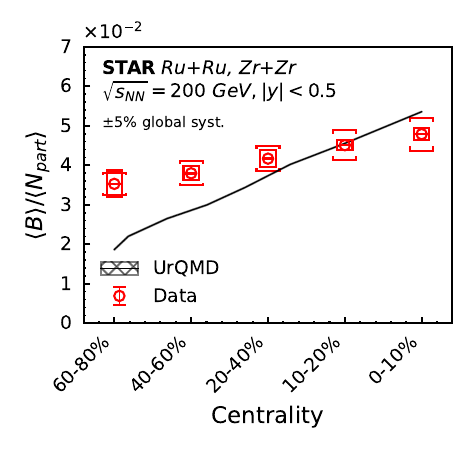}}
\end{minipage}
\caption[]{{\bf Comparison between data and UrQMD.} Distributions of net-charge difference (left) and average net-baryon yield (right), scaled by \npart, between \RuRu\ and \ZrZr\ collisions at \sqrtsNN\ = 200 GeV as a function of centrality. Red circles represent data with the boxes (vertical bars) around them representing systematic (statistical) uncertainties. Uncertainties due to feed-down correction for net-baryon yields are shown separately as brackets. UrQMD calculations~\cite{Lv:2023fxk} are shown as hatched bands for comparison. In the right panel, the hatched band collapses to a curve due to very small statistical uncertainties. }
\label{fig:Isobar_B_Q}
\end{figure}

The net-charge difference between \RuRu\ and \ZrZr\ collisions ($\Delta Q$) and the average net-baryon yield ($\langle B\rangle$) among the two collision systems are shown as a function of centrality \cite{STAR:2021mii} in the left and right panels of Fig. \ref{fig:Isobar_B_Q}, respectively. Additional uncertainties arising from unmeasured nuclei in the baryon number calculation, different extrapolation functions used for invariant yields and double ratios, and feed-down related parameters, are also included. Both quantities are scaled with \npart\ and compared to UrQMD calculations \cite{Bass:1998ca,Bleicher:1999xi}, which assign valence quarks as the baryon number carriers. For the baryon transport, the UrQMD model employs a Gaussian parameterization for the longitudinal fragmentation to produce more baryons than the Field-Feynman fragmentation~\cite{Bleicher:1999xi}. Such a parameterization was tuned to match previous experimental data on proton production in heavy-ion collisions by enhancing quark stopping at mid-rapidity, and thus provides a good description of the baryon transport in central \RuRu\ and \ZrZr\ collisions (right panel of Fig. \ref{fig:Isobar_B_Q}). However, this tuning results in a factor of 3 larger charge transport difference between \RuRu\ and \ZrZr\ collisions from UrQMD than that in data (left panel of Fig. \ref{fig:Isobar_B_Q}).

For the measurement of $\langle B\rangle/\Delta Q$ ratio, the dominant uncertainties are those associated with feed-down related parameters (6.8-7.2\%) and extrapolations of particle yields (2.0-8.3\%) and double ratios (3.2-4.6\%), where the uncertainty ranges result from their centrality dependence. Other sources of uncertainties typically contribute less than 2\%.  